\documentclass[conference]{IEEEtran}
\IEEEoverridecommandlockouts
\usepackage{cite}
\usepackage{amsmath,amssymb,amsfonts}
\usepackage{algorithmic}
\usepackage{graphicx}
\usepackage{textcomp}
\usepackage{xcolor}
\usepackage{booktabs} 
\usepackage{paralist}
\usepackage{multirow}
\usepackage{makecell}

\def\BibTeX{{\rm B\kern-.05em{\sc i\kern-.025em b}\kern-.08em
    T\kern-.1667em\lower.7ex\hbox{E}\kern-.125emX}}
\begin{document}

\title{Generating Pixel Art Character Sprites using GANs}

\author{
    \IEEEauthorblockN{Flávio Coutinho\IEEEauthorrefmark{1}\IEEEauthorrefmark{2} and Luiz Chaimowicz\IEEEauthorrefmark{1}}
    
    \IEEEauthorblockA{\IEEEauthorrefmark{1}Departamento de Ciência da Computação, Universidade Federal de Minas Gerais, 
    Belo Horizonte, Brazil \\
    Email: \{flavioro, chaimo\}@dcc.ufmg.br}

    \IEEEauthorblockA{\IEEEauthorrefmark{2}Departamento de Computação,
    Centro Federal de Educação Tecnológica de Minas Gerais, 
    Belo Horizonte, Brazil \\
    Email: fegemo@cefetmg.br}
}
    


\maketitle

\begin{abstract}
Iterating on creating pixel art character sprite sheets is essential to the game development process. However, it can take a lot of effort until the final versions containing different poses and animation clips are achieved. This paper investigates using conditional generative adversarial networks to aid the designers in creating such sprite sheets. We propose an architecture based on Pix2Pix to generate images of characters facing a target side (e.g., right) given sprites of them in a source pose (e.g., front). Experiments with small pixel art datasets yielded promising results, resulting in models with varying degrees of generalization, sometimes capable of generating images very close to the ground truth. We analyze the results through visual inspection and quantitatively with FID.
\end{abstract}


\section{Introduction}
\label{introduction}

Procedural content generation (PCG) can be described as the \textit{``algorithmic creation of game content with limited or indirect user input''} (p.~6) \cite{Togelius2011WhatBorderline}. It can be used either online in games or offline during development. In the former case, it is usually used to generate game levels to enable richer replay value, as one playthrough can be very different from the other. In the latter case, the users can be the game developers: the level designers receiving aid to generate content or designers when creating art assets.

In the case of online PCG, a common criticism is that such systems usually create repetitive content, making it less memorable to players \cite{Karavolos2015}. To some extent, PCG tries to replace human designers in their activity by developing programs that create content, and programmers do not necessarily have good game and level design skills. However, when offline methods are embedded in game design tools, for example, PCG can \textit{``augment the creativity of individual human creators''} (p.~3)~\cite{Togelius2016Introduction} instead of trying to replace them. As an illustration, some PCG systems let human creators and algorithms work together to produce game content. The automation of only part of the content creation pipeline is usually designated as mixed-initiative PCG, when both the algorithm and the human creator can interact with the content being generated in a co-creation process~\cite{Smith2011Tanagra:Design,Baldwin2017TowardsGeneration,Liapis2016Mixed-initiativeCreation}.

In this work, we present an architecture for generating pixel art character sprites in a target pose given its image in a source direction: for example, our approach can create an image of a character facing to the right from another one of it facing front. Such a model can aid designers during their creation process when integrated with a PCG tool.

The proposed architecture is based on Pix2Pix~\cite{Isola2017Image-to-ImageNetworks}. We trained and experimented with models using different datasets. The generated images were evaluated through visual inspection and calculating the Fréchet Inception Distance \cite{Heusel2017GANsEquilibrium} between the model's output and the ground truth for a more quantitative assessment.

The four datasets were small and had a number of training examples varying from 184 images to 776 in the largest one. As a result, in a dataset in which characters are modularly built by assembling different parts of the body, the model generated images with high perceptual quality, with almost no divergence from the ground truth. In the other datasets, the results had different quality levels, with some sprites suffering from high-frequency color noise. On average, the characters' shapes were more similar to the target images than their colors. Overall, the generated shape and the colors still resembled the ground truth, even for some lesser quality results.

\section{Related Work}
Serpa and Rodrigues \cite{Serpa2019TowardsSheets} tackled the generation of two types of pixel art sprites of game characters from line art sketches: (a) a grayscale sprite encoding shading information and (b) a colored sprite with body-part segmentation. The grayscale sprite can have 6 shades of gray and encodes how the image responds to lighting. On the other hand, the colored sprite segments pieces of the character (head, face, hair, limbs, trunk, etc.) by assigning one of 42 colors to each region.

The authors adapted the Pix2Pix architecture \cite{Isola2017Image-to-ImageNetworks} to generate two sprites from the same input. It comprises a U-net generator and a patch-based discriminator. The modifications include:
\begin{inparaenum}[1.]
    \item the U-net generator has two decoders, one for each sprite (gray and colored);
    \item the use of an $L_2$ loss function instead of $L_1$ used in Pix2Pix; and
    \item ELU activation functions instead of the leaky ReLU function.
\end{inparaenum}

In their experiments, the authors trained the model separately for the sprites of two characters (85 and 530 training examples). The generated grayscale images were very close to the ground truth in both characters. For the color sprites (segmentation of body parts), the model trained for the character with larger region segments (it had large arms and torso) generated better results for input sprites with similar poses in the training set. But for unseen poses, the generated images contained a lot of high-frequency noise.

Jiang and Sweetser \cite{Jiang2021GAN-AssistedGeneration} proposed a generative model for automatic coloring of game sprites, turning single-channel images into colored ones. It is also based on the Pix2Pix architecture \cite{Isola2017Image-to-ImageNetworks} and processes images using the YUV representation rather than in the RGB color space.

Experiments showed that the model trained with YUV inputs yielded better-colored images than when represented with RGB. Such improvement is attributed to the fact that when working in YUV, the network had to learn only the U and V channels, using Y as a redundancy of the condition.

Gonzalez, Guzdial, and Ramos \cite{Gonzalez2020GeneratingLearning} approached the problem of translating (Pokémon) pixel art characters from a source domain (Pokémon type, e.g., fire) into a version of the same sprite but in a target domain (e.g., grass). They used a Variational Autoencoder (VAE) with convolutional layers.

With a small dataset composed of 974 images, the authors decided to experiment with pre-training the model with the Anime Face Dataset\footnote{AFD: https://www.kaggle.com/splcher/animefacedataset} (AFD). The reconstruction of the original images yielded blurry results with shapes faintly resembling the original but with noisy colors.

The domain transfer (type swap) task had more shape degradation, usually with high-frequency noise and dangling pixels outside the intended character shape. Color-wise, using transfer learning provided images with colors that resembled the ones from the target domain (i.e., Pokémon type).

Hong, Kim, and Kang \cite{Hong2019GameGAN} proposed a system that takes two sprites, one representing a character's shape and color and the other comprising a target pose on which to draw the first. The authors used a multiple discriminator GAN (MDGAN).

The generator takes as input an image with the target pose information (bone graph) and another with shape and color, and it uses separate encoders for each. Its goal is to generate a sprite of the character with the shape and color of the second input image but in the pose represented in the first. There are two discriminators, the first being responsible for determining whether two images share the same color and shape, and the second extracting the pose of a character sprite.

The authors compared their approach with an adapted version of Pix2Pix, among other architectures, and MDGAN had the best outcome. However, albeit successful in the proposed experiments, the system requires a bone graph dataset matching characters' positions in the shape and color dataset. For that reason, the authors artificially tailored datasets with non-pixel art images, and using such an approach would make it difficult with real pixel art character sprites.

Our work also adapts Pix2Pix \cite{Shaker2016ProceduralGames,Jiang2021GAN-AssistedGeneration} to generate pixel art character sprites. But differently from the related works, we want to translate characters drawn on a source to a target side. Regarding the particular task being tackled, MDGAN \cite{Hong2019GameGAN} is the most similar work. Still, it requires a second large dataset composed of pose information (bone graph), which is typically unavailable, especially for pixel art. Regarding the image representation, we consider RGBA channels versus RGB in \cite{Serpa2019TowardsSheets,Hong2019GameGAN}, YUV in \cite{Jiang2021GAN-AssistedGeneration}, and HSV in \cite{Gonzalez2020GeneratingLearning}. We also study the impact of the alpha channel in one of our experiments.

\section{Architecture Overview}
\begin{figure*}[htb]
    \centering
    \includegraphics[width=1.\linewidth,interpolate=false]{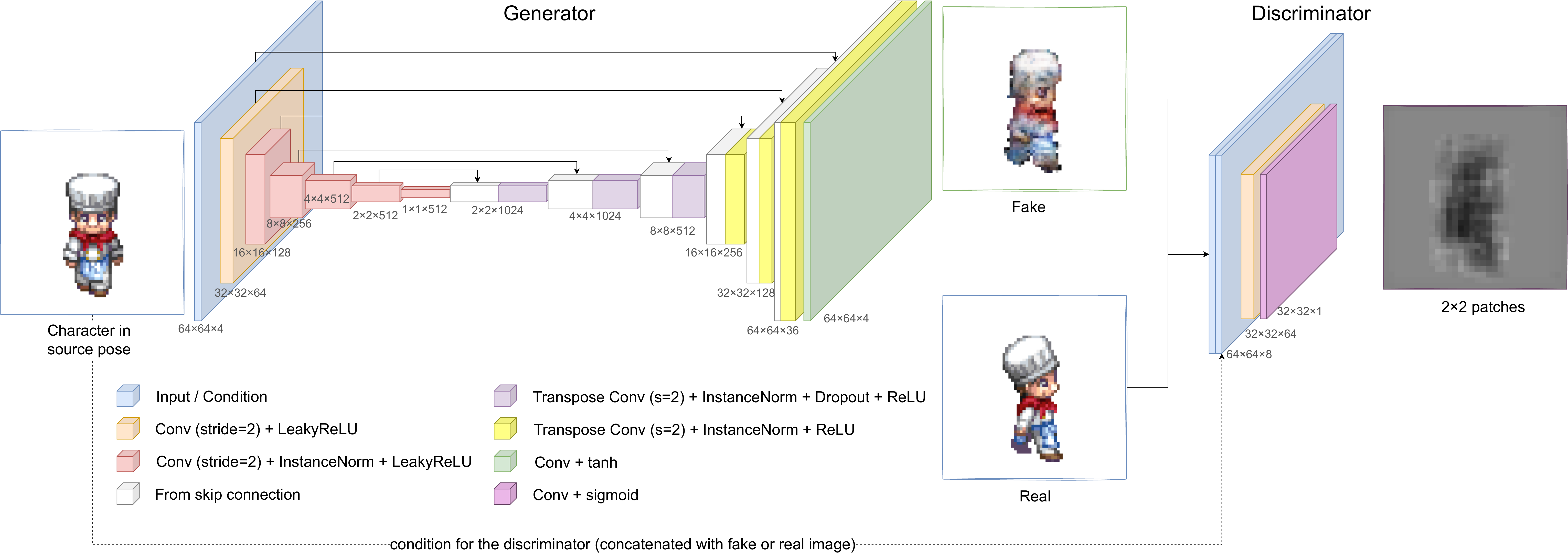}  
    \caption{Generator (left) and discriminator (right) network architecture.}
    \label{fig:architecture}
\end{figure*}

Similar to some related works \cite{Serpa2019TowardsSheets, Jiang2021GAN-AssistedGeneration}, our architecture is based on Pix2Pix. Each model we train can translate images of a dataset from a source into a target pose. Next, we describe our generator and discriminator networks. As presented in \figurename~\ref{fig:architecture}, the input to the generator is a 64$\times$64$\times$4 image with RGBA channels in the [-1, 1] domain.

\subsection{Generator}
The generator is a U-net with 64$\times$64$\times$4 input and output. The first layers downscale the image to 1$\times$1, and the last upscale the data to its original resolution.

Each downsampling step is composed of a convolution (4$\times$4 kernels), instance normalization (except for the first one), and leaky ReLU layers, and it reduces the spatial resolution by half. Considering the input has 64$\times$64 pixels, there are 6 downsampling blocks.

The decoder is a reflection analogy of the encoder, with upscaling steps instead. Each block comprises a transpose convolution (4$\times$4 kernels), instance normalization, an optional dropout (50\%), and ReLU activation layers. The first three blocks have dropout regularization, and the number of filters is the same as in the encoder but in reverse order. The last upsampling block has a $\tanh$ activation, so it outputs pixel intensities in the range of [-1, 1]. To allow the network to learn custom downsampling/upsampling mechanisms, it changes the resolution only through fractionally/strided convolutions \cite{Radford2015UnsupervisedNetworks}. There are skip connections from the output of the $i^{th}$ encoder layer to the $n^{th}-i$ decoder one, with $n = 6$, to preserve spatial information.

The loss function for the generator is similar to the one in \cite{Isola2017Image-to-ImageNetworks}, but uses a non-saturating adversarial part. We also use the ${L_1}$ distance between the real and generated images, with the hyperparameter $\lambda$ set to 100. The generator's loss is:
\begin{equation}
	\mathcal{L}_G=-\mathbb{E}_{x}[\log{D(G(x))}]+\lambda\mathbb{E}_{x,y}[||y-G(x)||_1]
\end{equation}
where $x$ represents images in source pose, and $y$ in the target; and $||\cdot||_1$ is the $L_1$ distance (absolute value of the differences of the patches from generated to target images).

\subsection{Discriminator}
Our discriminator is a PatchGAN \cite{Isola2017Image-to-ImageNetworks}: a conditional classifier that takes a source image (e.g., character facing front) as a condition, and either a real or a generated sprite in the target direction (e.g., facing right), splits it into same-size square patches, and discriminates each region as being real or fake.

The rationale behind splitting the output in patches is to enable local discrimination instead of providing a single global value of a sprite being either real or fake. Doing so allows penalizing more parts of the images that require more work from the generator. As noted by the architecture authors, while the generator's $L_1$ loss term steers the generation towards the target images (but leads to blurry results), the patch discrimination works as a texture loss.

\begin{figure}[htb]
    \centering
    \includegraphics[width=0.75\linewidth,interpolate=false]{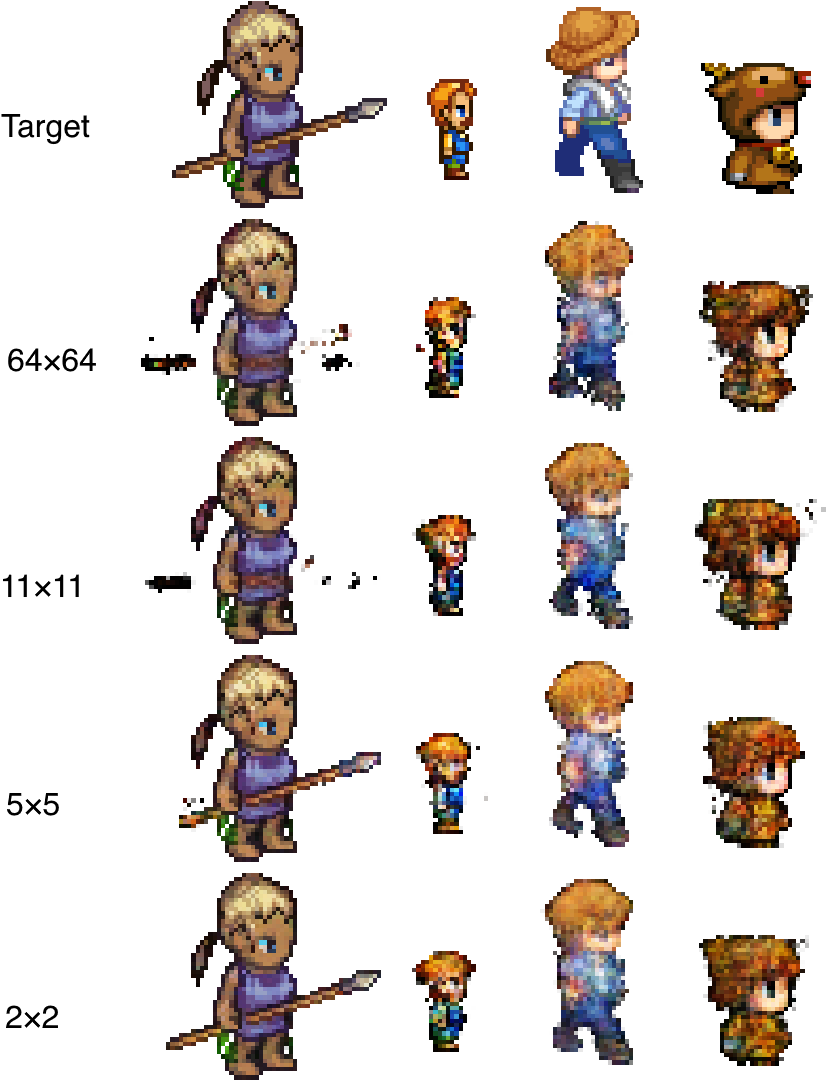}  
    \caption{Outputs of models with varying patch sizes for the discriminator.}
    \label{fig:patch-study}
\end{figure}

We experimented with different patch sizes: 2$\times$2, 5$\times$5, 11$\times$11, and 64$\times$64 (single patch) and \figurename~\ref{fig:patch-study} shows a comparison. The patch size of 2$\times$2 achieved the closest result to the ground truth. The other images presented some color and shape noise, with the models using larger patches yielding worse results. In particular, the model with a single patch suffered from dangling pixels outside the sprite shape.

We attribute the better results with smaller patches to the discriminator being penalized by misclassifying 2$\times$2 regions individually. As pixel art sprites typically have low resolution, each pixel carries a lot of information and should consider mostly its local vicinity. In such a setting, there are very small regions of low frequency (same color or just a slight variation). Hence, smaller patches evaluate not only the texture of the area but also the shape edges. Furthermore, it may be due to that double responsibility that the images present high color variation even in parts that should have a single or a few colors. For such reasons, we chose to use 2$\times$2 patches. Ultimately, the resulting shape and colors in all patch sizes resemble the ground truth to some degree.

Regarding its layers, the discriminator is composed of the same downsampling blocks used by the generator (\figurename~\ref{fig:architecture}). Moreover, its loss function is the same as for conditional GANs~\cite{Mirza2014ConditionalNets}, which can be calculated as a binary cross-entropy between the real images it discriminated as fake and the fake discriminated as real. The only change is that because the discriminator's output is not a single number per image but one for each patch, it is first reduced to the mean value.

\section{Experiments}
\label{sec:experiments}

We conducted different experiments to evaluate the model architecture. First, we investigated how well the model performed on the test data on the individual datasets.
Then, the results were analyzed through visual inspection and quantitatively using the Fréchet Inception Distance (FID)~\cite{Heusel2017GANsEquilibrium}.

During the analysis, we also investigated the generalization capacity of the models under different situations:
\begin{inparaenum}[(a)]
\item changing the colors of either one part or
\item the majority of a sprite, and
\item slightly changing the character pose.
\end{inparaenum}

Last, we evaluate whether the model provides better quality results when considering the alpha channel of the images or using only the RGB components.

\subsection{Datasets}
\begin{figure}[htb]
    \centering
    \includegraphics[width=.7\linewidth,interpolate=false]{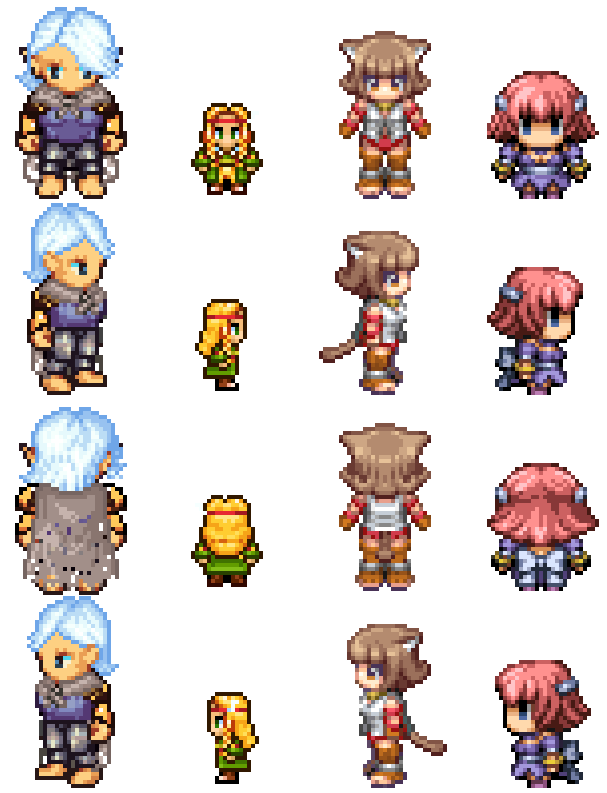}  
    \caption{Sample images from the dataset showing 4 different sizes/art styles (columns) in 4 directions (rows).}
    \label{fig:dataset}
\end{figure}

We assembled a dataset of pixel art character sprites from different sources. It contemplates characters in four directions -- front, right, back, and left -- and comprises four different sizes and art styles (\figurename~\ref{fig:dataset}).

The four image sources had different character sizes, so the smaller ones were transparency-padded to the largest size, which was 64$\times$64. Also, we created an alpha channel with the character shape for the images that lacked one. Table~\ref{tab:dataset} presents the sizes and number of images per character pose. Some sources had 3 frames for each character side, as they were extracted from sprite sheets with walking animations.

\begin{table}[htb]
\caption{Description of the datasets.}
\label{tab:dataset}
\begin{center}
\begin{small}
\begin{sc}
\begin{tabular}{@{}lcrr@{}}
\toprule
Source & Size & Examples & Per pose \\
\midrule
Tiny Hero           & 64$\times$64$\times$4   & 912   & 1 \\
RPG Maker 2000      & 32$\times$24$\times$3   & 216   & 3 \\
RPG Maker XP        & 48$\times$32$\times$4   & 294   & 3 \\
RPG Maker VX Ace    & 32$\times$32$\times$4   & 408   & 3 \\
\bottomrule
\end{tabular}
\end{sc}
\end{small}
\end{center}
\end{table}

The Tiny Hero sprites were modularly created by assembling previously-drawn body pieces to form each character. For this reason, it is usual for different characters to present the same or similar shapes, but with a different color or a different combination of parts.

\subsection{Training Procedure}
Training used an 85\% split. The model received images in batches of 1 sample. The generator's and discriminator's weights were optimized with Adam (momentum of $\beta_1=0.5$ and $\beta_2=0.999$) using a fixed learning rate of $0.0002$. Instead of running the optimization for some epochs, as the batch contained a single image, the process was split into training steps. All of the models trained for 40,000 steps.

We trained the models on a hardware with an Intel Core i7 7700HQ and an NVidia GTX 1050 Ti GPU and it took an average of 01h30m.

\section{Results}
We now present the generated images in different settings, analyze the results through visual inspection, and then quantitatively compare them using FID\footnote{Note on FID: the distances should be calculated with thousands of images~\cite{Heusel2017GANsEquilibrium}. However, the metric still applies even with our tiny datasets.}.

\subsection{Individual Datasets}
As each dataset had a different number of training examples and the training steps were fixed, each model trained for a different number of epochs (see Table~\ref{tab:experiment-dataset-summary}).

In all experiments, the models were trained to translate a front-facing character into the pose facing right. \figurename~\ref{fig:result-individual-one} and \ref{fig:result-individual-two} show results of the models trained with the individual datasets. All examples were hand-picked from the test data and organized vertically with better results at the top.

All training images could be reproduced with high fidelity by the models with all datasets, which is also indicated by their low FID values. Next, we analyze the models considering only images from the test.

\begin{table}[htb]
\caption{Training size, epochs and FID per dataset (40k steps).}
\label{tab:experiment-dataset-summary}
\begin{center}
\begin{small}
\begin{sc}
\begin{tabular}{@{}lrrrr@{}}
\toprule
\multirow{2}{*}{Dataset} &
\multirow{2}{*}{\makecell{Train\\size}} &
\multirow{2}{*}{Epochs} &
\multicolumn{2}{c}{FID} \\
& & & train & test \\
\midrule
Tiny Hero           & 776           & $\approx$52   & 0.092 & 0.115 \\
RPG Maker 2000      & 184           & $\approx$217  & 0.091 & 2.306 \\
RPG Maker XP        & 250           & 160           & 0.264 & 9.493 \\
RPG Maker VX Ace    & 347           & $\approx$115  & 0.263 & 5.495 \\
\bottomrule
\end{tabular}
\end{sc}
\end{small}
\end{center}
\end{table}

\subsubsection{Tiny Hero sprites}
\label{sec:tiny-hero}

\begin{figure}[htb]
    \centering    \includegraphics[width=1.0\linewidth,interpolate=false]{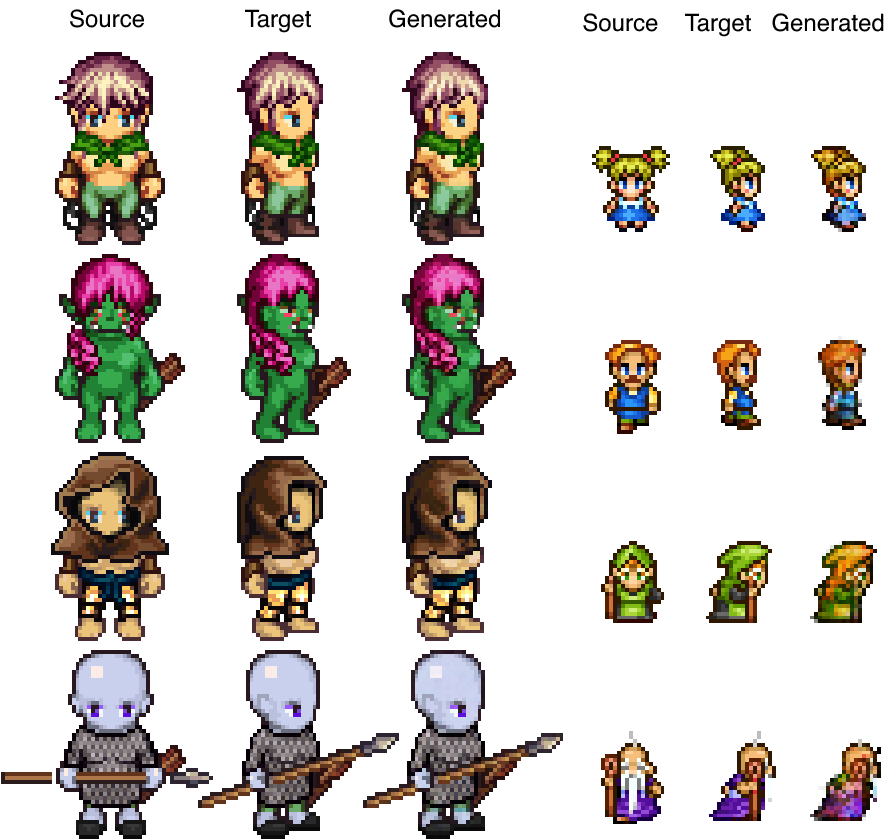}  
    \caption{Test examples from \textit{Tiny Hero} (left) and \textit{RPG Maker 2000} (right).}
    \label{fig:result-individual-one}
\end{figure}

The images generated for Tiny Hero (\figurename~\ref{fig:result-individual-one}, left) were perceptually identical to the ground truth, which is corroborated by the FID score of 0.115. Although it looks rather impressive at first, such a nice result could only be achieved due to how the Tiny Hero dataset was built: the characters were created by assembling body parts and accessories. So although the model had not seen the test images nor their ground truth, it learned how to translate each body part while training. This test does not show that the proposed model can generalize the translation but can memorize segments of the sprites it sees during training.

\subsubsection{RPG Maker 2000}
The first row of \figurename~\ref{fig:result-individual-one} (right) shows a child girl translated with only minor color differences to the ground truth. Although that sprite was not on the training set, there was another one with the exact shape but different hair and dress color. In this case, the model could understand their shapes and translate only the colors of parts of a sprite. Something similar happened to the granny and gramps in the third and fourth rows -- the model learned a sprite with the same shape but wearing an orange dress. In this case, it could not fully translate the colors, but it preserved the shape for the old lady and partially preserved it for the graybeard man.

\begin{figure}[htb]
    \centering
    \includegraphics[width=0.85\linewidth,interpolate=false]{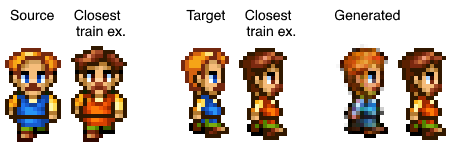}  
    \caption{Front and right view of a test sprite (blue clothes) with the closest example (orange) in the training set from \textit{RPG Maker 2000}.}
    \label{fig:result-generalization-rpgmaker2000}
\end{figure}

The man in the second row of \figurename~\ref{fig:result-individual-one} was an interesting result suggesting a more profound generalization, as its translation uses information of a sprite with similar, but not equal shape, seen during training, but using different colors and slightly different shape. \figurename~\ref{fig:result-generalization-rpgmaker2000} shows the test image input, ground truth, and generated image, with the closest match from the training set to the side for comparison. We can note that the front-facing characters look similar, with differences in colors, the size and position of the hair, and the design of the mustache. The generated image contains some noise, and the edges inside the shape are not crisp, but it did not reuse the same hair and mustache shape as the other character.

The FID value for the generated images in the dataset was 2.306 (Table~\ref{tab:experiment-dataset-summary}) and, after Tiny Hero, it was the smallest value, indicating the generated images were close to the ground truth.

\subsubsection{RPG Maker XP}
\begin{figure}[htb]
    \centering
    \includegraphics[width=1.0\linewidth,interpolate=false]{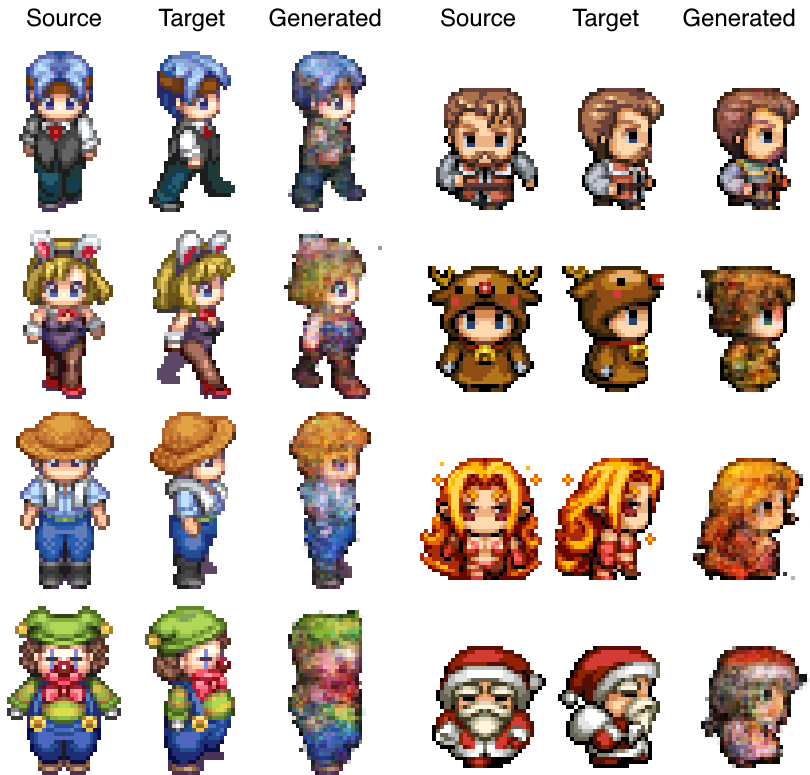}  
    \caption{Test examples from \textit{RPG Maker XP} (left) and \textit{RPGM VX Ace} (right).}
    \label{fig:result-individual-two}
\end{figure}

The model trained with the RPG Maker XP sprites had the highest FID score (9.493), indicating lower translation quality. \figurename~\ref{fig:result-individual-two} (left) shows examples of the best quality translations (top) to the worst ones (bottom). Such sprites are larger than those from RPG Maker 2000 (32$\times$24 vs. 48$\times$32) and not modular as in Tiny Hero. Furthermore, there are no sprites with the same shape with just some color variation, making it a suitable lower bound for the translation quality of the proposed model.

The image generated in the first row of \figurename~\ref{fig:result-individual-two} (left) had higher quality than the others. That happened because this dataset has 3 images per character pose, which are frames of a walking cycle. In this case, the frame of the character standing still stayed in the training set, with the other two as part of the test. Although the generated image contains some noise, it is possible to see the faint color of the ribbon in his head, and the overall shape is correct. This suggests that small changes in the pose are at least partially generalizable.

Each character in the middle rows (dealer girl and farmboy) has distinct features -- a bunny tiara and a straw hat. Unfortunately, the model could not correctly translate such features' shapes, but we can notice that some traits, such as the most prominent colors and prominences, exist.  

Last, the generated image for the clown in the fourth row has a lot of high-frequency noise. Again, the colors appear in meaningful positions, but there is very little information to separate them visually.

\subsubsection{RPG Maker VX Ace}
The first row of \figurename~\ref{fig:result-individual-two} (right) presents the best result, and the character is a color variation of another sprite seen during training. Except for the color of the left hand, there is very little perceptual difference with the ground truth.

The characters in the middle rows have distinctive features, such as in the RPG Maker XP test. They differ from the ground truth, but the sprites have the overall colors and an approximate shape. There was high-frequency color noise and blurry edges inside the shape. In the fourth row, Santa Claus was generated implausibly and, even worse, without its sack of gifts.

The comparison of the FID between the model output and the ground truth images was 5.495 -- a value between RPG Maker 2000 and RPG Maker XP.

\subsection{Alpha Channel}
We experimented with training the model to receive, process, and output images with the RGBA components and with only RGB. In character sprites, the images typically have either an alpha channel or a background key color which is ignored by the game engine when rendering.

\begin{figure}[htb]
    \centering
    \includegraphics[width=0.75\linewidth,interpolate=false]{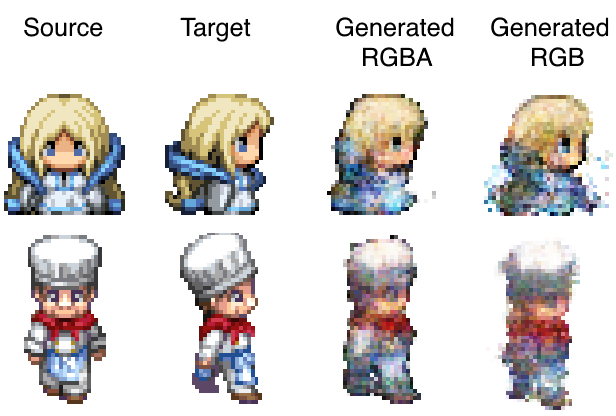}  
    \caption{Comparison of model outputs when training with RGBA vs. RGB.}
    \label{fig:alpha-study}
\end{figure}

Although the alpha channel is mainly used as a boolean mask (so is the case in all the datasets used in this work), and a color key would also work, we found that our GAN models do use the redundant information in the channel, allowing it to generate shapes with higher fidelity to the ground truth and avoiding dangling pixels outside the expected shape region. \figurename~\ref{fig:alpha-study} shows a comparison of images generated by models using RGBA and with RGB only.

\section{Final Remarks}

This work proposes an architecture based on Pix2Pix to create models that can translate images of character sprites from a source to a target pose (e.g., front to right-facing). We studied which patch sizes for the discriminator yielded results with better quality and also found that using the information in the alpha channel improves the quality of the shapes.

We showed through experiments with different datasets that the model can translate the color of partial parts of the characters and its entirety -- although in some situations, the color transfer was not entirely correct. The model also indicated some generalization capacity to leverage the information of a character in a pose (front-facing, standing still) to satisfactorily translate a slightly different pose (front-facing, one foot forward).

Still, in the absence of more examples, the translations, in some cases, had low quality. Overall, when the model does not generalize well, the generated images suffer from having high-frequency color noise and very faint inner edges, resulting in unusable images.

For future work, one can experiment using the multi-domain architecture of StarGAN \cite{Choi2018StarGAN:Translation}, as it would allow a single model to learn all the mapping among all available poses instead of having one model for each. Another further investigation is to increase the model's generalizability by experimenting with different loss functions, network architectures, and image representations.

\bibliographystyle{IEEEtran}
\bibliography{IEEEabrv,main}

\begin{thebibliography}{10}
\providecommand{\url}[1]{#1}
\csname url@samestyle\endcsname
\providecommand{\newblock}{\relax}
\providecommand{\bibinfo}[2]{#2}
\providecommand{\BIBentrySTDinterwordspacing}{\spaceskip=0pt\relax}
\providecommand{\BIBentryALTinterwordstretchfactor}{4}
\providecommand{\BIBentryALTinterwordspacing}{\spaceskip=\fontdimen2\font plus
\BIBentryALTinterwordstretchfactor\fontdimen3\font minus
  \fontdimen4\font\relax}
\providecommand{\BIBforeignlanguage}[2]{{%
\expandafter\ifx\csname l@#1\endcsname\relax
\typeout{** WARNING: IEEEtran.bst: No hyphenation pattern has been}%
\typeout{** loaded for the language `#1'. Using the pattern for}%
\typeout{** the default language instead.}%
\else
\language=\csname l@#1\endcsname
\fi
#2}}
\providecommand{\BIBdecl}{\relax}
\BIBdecl

\bibitem{Togelius2011WhatBorderline}
\BIBentryALTinterwordspacing
J.~Togelius, E.~Kastbjerg, D.~Schedl, and G.~N. Yannakakis, ``{What is
  procedural content generation? Mario on the borderline},'' in \emph{ACM
  International Conference Proceeding Series}.\hskip 1em plus 0.5em minus
  0.4em\relax New York, New York, USA: ACM Press, 2011, pp. 1--6. [Online].
  Available: \url{http://portal.acm.org/citation.cfm?doid=2000919.2000922}
\BIBentrySTDinterwordspacing

\bibitem{Karavolos2015}
\BIBentryALTinterwordspacing
D.~Karavolos, A.~Bouwer, and R.~Bidarra, ``{Mixed-Initiative Design of Game
  Levels: Integrating Mission and Space into Level Generation},'' in
  \emph{Proceedings of the 10th International Conference on the Foundations of
  Digital Games}, 2015. [Online]. Available: \url{http://unity3d.com/}
\BIBentrySTDinterwordspacing

\bibitem{Togelius2016Introduction}
J.~Togelius, N.~Shaker, and M.~J. Nelson, ``{Introduction},'' in
  \emph{Procedural Content Generation in Games: A Textbook and an Overview of
  Current Research}.\hskip 1em plus 0.5em minus 0.4em\relax Springer, 2016,
  ch.~1.

\bibitem{Smith2011Tanagra:Design}
\BIBentryALTinterwordspacing
G.~Smith, J.~Whitehead, and M.~Mateas, ``{Tanagra: Reactive planning and
  constraint solving for mixed-initiative level design},'' in \emph{IEEE
  Transactions on Computational Intelligence and AI in Games}, vol.~3, no.~3, 9
  2011, pp. 201--215. [Online]. Available:
  \url{http://ieeexplore.ieee.org/document/5887401/}
\BIBentrySTDinterwordspacing

\bibitem{Baldwin2017TowardsGeneration}
\BIBentryALTinterwordspacing
A.~Baldwin, S.~Dahlskog, J.~M. Font, and J.~Holmberg, ``{Towards pattern-based
  mixed-initiative dungeon generation},'' in \emph{ACM International Conference
  Proceeding Series}, vol. Part F1301.\hskip 1em plus 0.5em minus 0.4em\relax
  New York, New York, USA: Association for Computing Machinery, 8 2017, pp.
  1--10. [Online]. Available:
  \url{http://dl.acm.org/citation.cfm?doid=3102071.3110572}
\BIBentrySTDinterwordspacing

\bibitem{Liapis2016Mixed-initiativeCreation}
A.~Liapis, G.~Smith, and N.~Shaker, ``{Mixed-initiative content creation},'' in
  \emph{Procedural Content Generation in Games: A Textbook and an Overview of
  Current Research}.\hskip 1em plus 0.5em minus 0.4em\relax Springer, 2016,
  ch.~11.

\bibitem{Isola2017Image-to-ImageNetworks}
\BIBentryALTinterwordspacing
P.~Isola, J.-Y. Zhu, T.~Zhou, and A.~A. Efros, ``{Image-to-Image Translation
  with Conditional Adversarial Networks},'' in \emph{2017 IEEE Conference on
  Computer Vision and Pattern Recognition (CVPR)}, vol. 2017-Janua.\hskip 1em
  plus 0.5em minus 0.4em\relax IEEE, 7 2017, pp. 5967--5976. [Online].
  Available: \url{http://ieeexplore.ieee.org/document/8100115/}
\BIBentrySTDinterwordspacing

\bibitem{Heusel2017GANsEquilibrium}
\BIBentryALTinterwordspacing
M.~Heusel, H.~Ramsauer, T.~Unterthiner, B.~Nessler, and S.~Hochreiter, ``{GANs
  Trained by a Two Time-Scale Update Rule Converge to a Local Nash
  Equilibrium},'' \emph{Advances in Neural Information Processing Systems},
  vol. 2017-December, pp. 6627--6638, 6 2017. [Online]. Available:
  \url{https://arxiv.org/abs/1706.08500v6}
\BIBentrySTDinterwordspacing

\bibitem{Serpa2019TowardsSheets}
\BIBentryALTinterwordspacing
Y.~R. Serpa and M.~A.~F. Rodrigues, ``{Towards Machine-Learning Assisted Asset
  Generation for Games: A Study on Pixel Art Sprite Sheets},'' in \emph{2019
  18th Brazilian Symposium on Computer Games and Digital Entertainment
  (SBGames)}, vol. 2019-Octob.\hskip 1em plus 0.5em minus 0.4em\relax Rio de
  Janeiro: IEEE, 10 2019, pp. 182--191. [Online]. Available:
  \url{https://ieeexplore.ieee.org/document/8924853/}
\BIBentrySTDinterwordspacing

\bibitem{Jiang2021GAN-AssistedGeneration}
Z.~Jiang and P.~Sweetser, ``{GAN-Assisted YUV Pixel Art Generation},'' in
  \emph{Australasian Joint Conference on Artificial Intelligence}, 2021, pp.
  1--12.

\bibitem{Gonzalez2020GeneratingLearning}
\BIBentryALTinterwordspacing
A.~Gonzalez, M.~Guzdial, and F.~Ramos, ``{Generating Gameplay-Relevant Art
  Assets with Transfer Learning},'' in \emph{Proceedings of the AIIDE Workshop
  on Experimental AI in Games}, 10 2020, pp. 1--7. [Online]. Available:
  \url{http://arxiv.org/abs/2010.01681}
\BIBentrySTDinterwordspacing

\bibitem{Hong2019GameGAN}
\BIBentryALTinterwordspacing
S.~Hong, S.~Kim, and S.~Kang, ``{Game sprite generator using a multi
  discriminator GAN},'' \emph{KSII Transactions on Internet and Information
  Systems}, vol.~13, no.~8, pp. 4255--4269, 2019. [Online]. Available:
  \url{http://itiis.org/digital-library/manuscript/2473}
\BIBentrySTDinterwordspacing

\bibitem{Shaker2016ProceduralGames}
\BIBentryALTinterwordspacing
N.~Shaker, J.~Togelius, and M.~J.~Nelson, \emph{{Procedural Content Generation
  in Games}}.\hskip 1em plus 0.5em minus 0.4em\relax Springer, 2016. [Online].
  Available: \url{http://pcgbook.com/}
\BIBentrySTDinterwordspacing

\bibitem{Radford2015UnsupervisedNetworks}
\BIBentryALTinterwordspacing
A.~Radford, L.~Metz, and S.~Chintala, ``{Unsupervised Representation Learning
  with Deep Convolutional Generative Adversarial Networks},'' \emph{4th
  International Conference on Learning Representations, ICLR 2016 - Conference
  Track Proceedings}, 11 2015. [Online]. Available:
  \url{http://arxiv.org/abs/1511.06434}
\BIBentrySTDinterwordspacing

\bibitem{Mirza2014ConditionalNets}
\BIBentryALTinterwordspacing
M.~Mirza and S.~Osindero, ``{Conditional Generative Adversarial Nets},''
  \emph{arXiv preprint arXiv:1411.1784}, 11 2014. [Online]. Available:
  \url{http://arxiv.org/abs/1411.1784}
\BIBentrySTDinterwordspacing

\bibitem{Choi2018StarGAN:Translation}
\BIBentryALTinterwordspacing
Y.~Choi, M.~Choi, M.~Kim, J.-W. Ha, S.~Kim, and J.~Choo, ``{StarGAN: Unified
  Generative Adversarial Networks for Multi-domain Image-to-Image
  Translation},'' in \emph{2018 IEEE/CVF Conference on Computer Vision and
  Pattern Recognition}.\hskip 1em plus 0.5em minus 0.4em\relax IEEE, 6 2018,
  pp. 8789--8797. [Online]. Available:
  \url{https://ieeexplore.ieee.org/document/8579014/}
\BIBentrySTDinterwordspacing

\end{thebibliography}

\end{document}